%% file: 0_Main.tex
\documentclass[runningheads]{llncs}

\usepackage[T1]{fontenc}

\usepackage{graphicx}
\usepackage{marvosym}
\usepackage{orcidlink}
\usepackage{booktabs}
\usepackage{multirow}
\usepackage{url}

\newcommand{\bslabel}[1]{\par\noindent\textbf{#1}}
\newcommand{\sig}{\rlap{$^{*}$}}

\hypersetup{
    colorlinks=true,
    linkcolor=black,
    citecolor=black,
    urlcolor=blue
}

\begin{document}

\title{Scientific Claim–Source Retrieval Revisited:\\ A Comparative Study of Style Transfer and Re-Ranking}

\author{Tobias Schreieder\inst{1}\textsuperscript{\Letter}\orcidlink{0009-0000-8268-4204} \and
Harsh Khandelwal\inst{2}\orcidlink{0009-0004-0237-5735} \and
Yu-Ling Zhong\inst{1}\orcidlink{0009-0007-2361-4944} \and
Michael Färber\inst{1}\orcidlink{0000-0001-5458-8645}}

\authorrunning{T. Schreieder et al.}
\titlerunning{Scientific Claim--Source Retrieval Revisited}

\institute{
TU Dresden \& ScaDS.AI Dresden/Leipzig, Dresden, Germany\\
\email{tobias.schreieder@tu-dresden.de}
\and
Friedrich-Alexander University of Erlangen--Nuremberg, Erlangen, Germany\\
}

\maketitle

\begin{abstract}
Scientific claims shared on social media are often difficult to verify and may contribute to the spread of misinformation. To address this challenge, automated fact verification systems require scientific claim--source retrieval, the task of identifying the source publication underlying a given claim. However, claims often differ substantially from their source publications in language, style, and specificity, making retrieval challenging. We present a comparative study of scientific claim--source retrieval on the CheckThat! 2026 benchmark across sparse and dense retrieval models. Our results show that translating claims into English outperforms both original and bilingual claim representations, while incorporating publication metadata provides additional retrieval gains by capturing indirect source references. In addition, we analyze four style transfer approaches and find that they improve retrieval performance for most models, although the optimal style depends on the underlying retrieval objective. Finally, we investigate similarity- and signal-based re-ranking approaches, introducing three novel re-ranking models based on attribution, entity overlap, and verification-based reasoning. Verification-based re-ranking yields additional gains beyond semantic similarity and achieves the best overall performance with an MRR@5 of 0.758.

\keywords{Scientific Claim--Source Retrieval \and Re-Ranking \and Text Style Transfer \and Multilinguality \and Large Language Models \and Fact Verification.}
\end{abstract}

\input{texts/1_Introduction}
\input{texts/2_Related-Work}
\input{texts/3_Experimental-Setup}
\input{texts/4_Methodology}
\input{texts/5_Evaluation}
\input{texts/6_Discussion}
\input{texts/7_Conclusion}

\section*{Acknowledgments}
The authors acknowledge the financial support by the Federal Ministry of Research, Technology and Space of Germany (BMFTR) and by the Saxon State Ministry for Science, Culture and Tourism in the programme Center of Excellence for AI Research „Center for Scalable Data Analytics and Artificial Intelligence Dresden/Leipzig“, project identification number: ScaDS.AI. 

Tobias Schreieder is supported by the BMFTR through a Software Campus project, project identification number: 16|S23070.

The authors also acknowledge computing resources provided by the NHR Center at TU Dresden, supported by the BMFTR and the participating state governments within the NHR framework.

The authors used AI-based assistance tools to support language editing, minor formatting, and coding tasks. These tools did not contribute to the intellectual content or scientific conclusions. All content was reviewed by the authors, who assume full responsibility for the publication.

\bibliographystyle{splncs04}
\bibliography{literature}

\end{document}

%% file: texts/1_Introduction.tex
\section{Introduction}
\label{sec:introduction}

Scientific communication has undergone a notable shift. While peer-reviewed journals and conferences remain central for communication within expert communities, social media platforms such as X (formerly Twitter) serve as additional channels that enable the rapid dissemination of research findings, the promotion of publications, and direct engagement with broader audiences~\cite{Guenther2023TwitterScienceCommunication,Haunschild2021TwitterOpioid}. Despite these benefits, social media also raises concerns about the credibility of scientific information. Scientific claims are often presented without sufficient context, expressed imprecisely, or only loosely connected to the underlying evidence, making their credibility difficult to assess. In the worst case, such claims may constitute misinformation, that is, information that contradicts current scientific evidence or consensus~\cite{SwireThompson2020Misinformation}, which is widespread on social media, particularly in areas such as vaccination, opioids, and noncommunicable diseases~\cite{Suarez-Lledo2021HealthMisinformation}.

Addressing these challenges requires automated fact verification systems that assess the factual correctness of claims with respect to relevant evidence. In this context, we focus on scientific claim--source retrieval, a key component of such systems that aims to identify the scientific publication that serves as the source for a given claim. Scientific claim--source retrieval has recently been introduced in the CheckThat! Lab and extended to multilingual settings~\cite{Alam2026CheckThat2025,Struss2026CheckThat2026}. However, this task is inherently difficult in the context of social media, where claims are often expressed in colloquial or simplified language. In addition, claims and their corresponding scientific publications may be expressed in different languages, introducing a challenge for retrieval systems given the predominantly English scientific literature. Furthermore, claims often contain only partial and implicit references to their sources, such as mentions of authors, institutions, or venues.

Our previous work on scientific claim--source retrieval demonstrated that zero-shot style transfer, in particular rewriting claims into a more formal style, generally improves retrieval performance~\cite{Schreieder2025Claim2Source}. While that study focused on monolingual style transfer, this work revisits style transfer on the multilingual CheckThat! 2026 benchmark by evaluating multiple large language models (LLMs). Beyond reproducing and extending the style-transfer analysis, we investigate re-ranking as a complementary retrieval strategy, enabling the first comparative study of style transfer and re-ranking for scientific claim--source retrieval. Code and resources are publicly available.\footnote{\url{https://github.com/faerber-lab/CheckThat2026}} Our contributions are as follows:

\begin{enumerate}
    \item We analyze multilingual claim representations and show the impact of original, translated, and bilingual claims on retrieval performance.
    
    \item We investigate the role of publication metadata and demonstrate its value for scientific claim--source retrieval.
    
    \item We extend prior work on zero-shot style transfer to a multilingual setting and provide a comparative analysis across sparse and dense retrieval models.
    
    \item We compare similarity-based re-rankers and propose three novel signal-based re-rankers that leverage attribution, entity overlap, and verification signals.
\end{enumerate}

%% file: texts/2_Related-Work.tex
\section{Related Work}
\label{sec:related_work}

The task of evidence retrieval aims to identify relevant sources for a given claim, often as part of claim verification. Scientific claim–source retrieval represents a specific instance of this problem, where the goal is to identify the referenced scientific paper. Early work relies on traditional retrieval and learning-to-rank methods~\cite{Ma2018TwitterRetrieval}, followed by a shift toward neural and transformer-based approaches such as BERT-based retrieval models~\cite{Soleimani2020BertRetrieval,Samarinas2021AutomatedFactChecking}. 
Hybrid retrieval has emerged as a strong paradigm that combines sparse and dense signals to leverage complementary strengths~\cite{Sundriyal2022Covid19Retrieval}. Recent work shows that hybrid approaches, especially when combined with re-ranking, consistently outperform single-model retrieval setups~\cite{Sager2025DeepRetrieval}.
To address heterogeneous sources, cross-genre retrieval frameworks were proposed to align scientific and journalistic evidence~\cite{Zuo2023CrossGenreRetrieval}. Data augmentation strategies that modify or reformulate claims have been investigated to improve retrieval performance~\cite{Schofield025DSGT}.
Subsequent work moved beyond relevance estimation by incorporating verifier feedback~\cite{Zhang2023FeedbackFactVerification} and modeling multi-step reasoning over interdependent evidence~\cite{Liao2023MUSER}. In addition, re-ranking strategies are critical, with listwise re-rankers outperforming pointwise approaches in the experiments on the CheckThat! 2025 dataset~\cite{Staudinger025ATOM}.
A growing line of work leverages LLMs to enhance retrieval and reasoning, including question decomposition~\cite{Chen2024ComplexClaimVerification}, contrastive re-ranking~\cite{Sriram2024ContrastiveLearning}, and sub-question enrichment strategies~\cite{Churina2024QuestionEnrichment}.
Finally, RAG frameworks have been used to jointly assess factual accuracy and relevance by generating evidence summaries from retrieved sources, which provides a more interpretable basis for claim verification~\cite{Upadhyay2025RAG}. LLM-based source filtering has been proposed to refine retrieved candidates~\cite{Besrour2025SQuAI}.
While prior work has advanced retrieval through dense methods, hybrid approaches, LLM-based reasoning, and style transfer, it largely relies on textual similarity signals. We extend this line of work by systematically comparing style transfer and introducing re-ranking signals based on attribution, entity overlap, and verification reasoning.

%% file: texts/3_Experimental-Setup.tex
\section{Experimental Setup}
\label{sec:experimental_setup}

This section introduces the dataset and retrieval models used in our study.

\subsection{The CheckThat! 2026 Dataset}
\label{sec:dataset}

We conduct our experiments on the CheckThat! 2026 \textit{Source Retrieval for Scientific Web Claims} dataset~\cite{Struss2026CheckThat2026}. The dataset consists of social media claims and their corresponding source publications. Claims are provided in English (EN), German (DE), and French (FR), while 10,000 English scientific publications form the candidate set. Since our study does not involve training, we restrict our experiments to the validation set, which contains 4,993 claims (3,905 EN, 386 DE, and 702 FR). The task is formulated as a known-item retrieval problem, where the goal is to retrieve the source publication corresponding to a claim from the candidate set. This is challenging because claims often refer to their sources only implicitly and may differ substantially in language and style.

\subsection{Retrieval Models}
\label{sec:retrieval_models}

We examine one sparse retrieval model and three dense models to provide a diverse set of retrieval paradigms for our experiments.

\bslabel{BM25.} \textit{BM25} is a widely used sparse retrieval model based on term frequency and inverse document frequency, relying on exact term matching to rank source documents~\cite{Robertson1994BM25,Robertson1994Okapi}. Its lexical nature makes it sensitive to claim formulation.

\bslabel{GTR.} Based on the T5 architecture, \textit{gtr-t5-xl} is an 11B-parameter dense model trained on large-scale datasets for complex retrieval tasks~\cite{Ni2022GTR}.

\bslabel{E5.} The dense model \textit{intfloat/e5-large-v2}, with 335M parameters, is based on a transformer architecture. It is trained with a contrastive objective on large-scale retrieval data for semantic retrieval tasks~\cite{Wang2024E5}.

\bslabel{GritLM.} \textit{GritLM/GritLM-7B} is a dense model with 7B parameters based on a decoder-only transformer architecture. It is trained with instruction tuning on large-scale corpora and performs well in zero-shot retrieval settings~\cite{Muennighoff2025GritLM}.

%% file: texts/4_Methodology.tex
\section{Facets of Scientific Claim–Source Retrieval}
\label{sec:methodology}

We conduct a comparative study of different approaches for scientific claim--source retrieval in multilingual settings. We analyze five key facets of the retrieval pipeline: (1) handling multilingual claims, (2) leveraging publication metadata, (3) query reformulation via style transfer, (4) similarity-based re-ranking, and (5) signal-based re-ranking. Each facet captures a distinct aspect of the problem, ranging from linguistic and stylistic variation in claims and incomplete or implicit source references to modeling choices in retrieval and re-ranking, while building on the best-performing configuration from preceding experiments.

\subsection{Handling Multilingual Claims}
\label{sec:methodology_multilinguality}

To address multilinguality, we evaluate three claim representations: original claims, translated claims, and bilingual claims. \textbf{Original claims} correspond to the claims in their respective languages (English, German, and French). For \textbf{translated claims}, we translate all claims into English using zero-shot translation with the LLM \textit{Qwen/Qwen3-8B} to analyze the impact of language on downstream retrieval performance. \textbf{Bilingual claims} are constructed by jointly encoding the original and translated claims, while for English claims only the original version is used. This setup allows us to compare original, translated, and bilingual claim representations for scientific claim--source retrieval.

\subsection{Leveraging Publication Metadata}
\label{sec:methodology_metadata}

As discussed in Section~\ref{sec:dataset}, claims may include only indirect references to their sources, for example through mentions of authors, institutions, or venues. To investigate whether such signals can improve retrieval performance, we compare two strategies for encoding sources.
In the \textbf{no metadata} setting, source representations consist only of the title and abstract provided in the dataset, which are concatenated into a single input sequence. In the \textbf{metadata} setting, we augment the source representation with additional structured information in the form of labeled fields: \textit{title}, \textit{abstract}, \textit{authors}, and \textit{venue}. To limit noise from long author lists, we include only the first and last three authors.
This comparison allows us to evaluate whether incorporating publication metadata improves retrieval performance in scenarios where claims refer to sources only implicitly.

\subsection{Style Transfer for Scientific Claims}
\label{sec:methodology_style_transfer}

\begin{figure}[t]
    \centering
    \includegraphics[width=\linewidth]{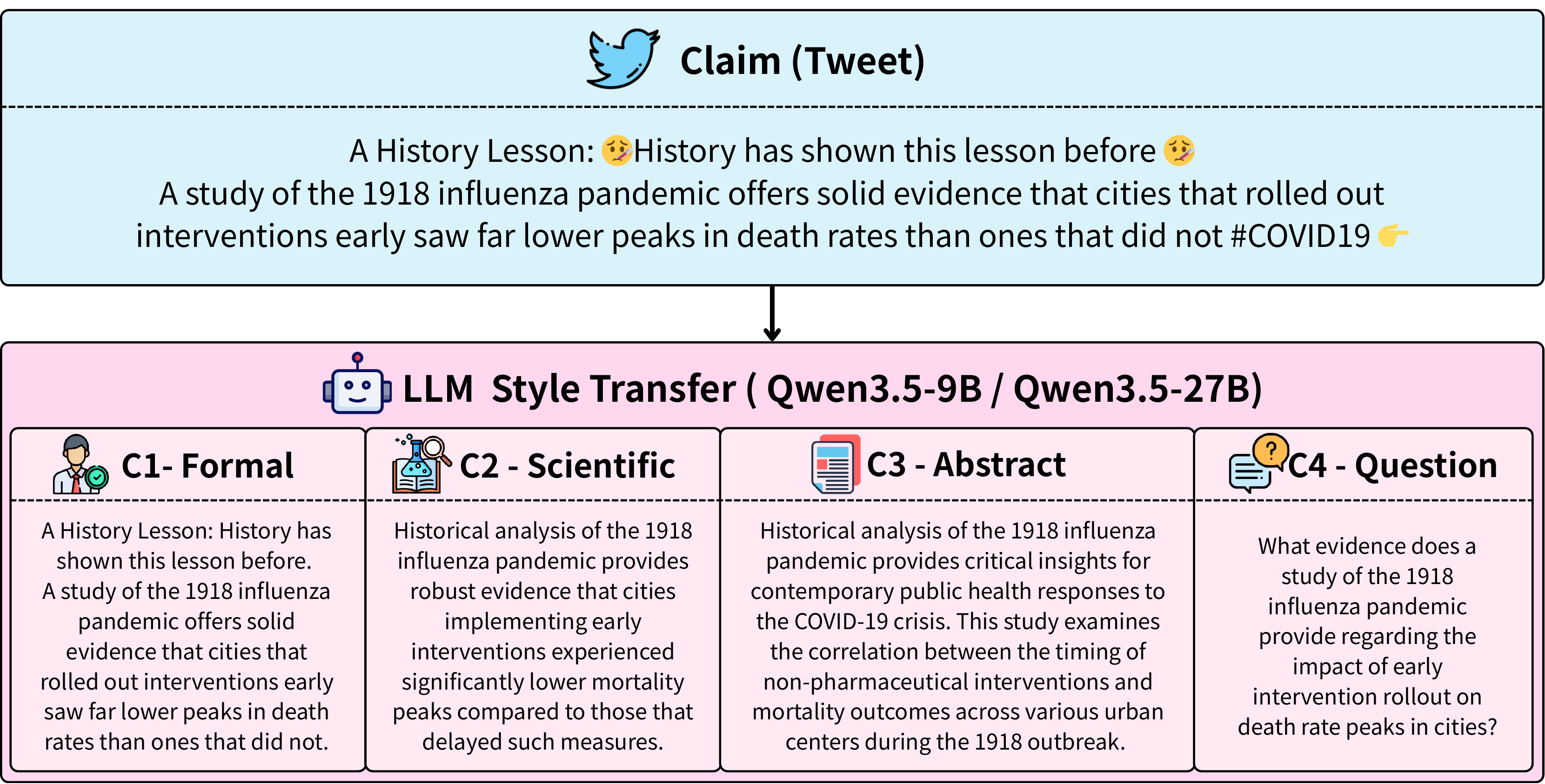}
    \caption{Illustration of the style transfer process. Given a claim, the LLM generates stylistic variants to create diverse representations while preserving the original meaning.}
    \label{fig:style_transfer}
\end{figure}

As illustrated in Figure~\ref{fig:style_transfer}, we apply style transfer to claims to improve their alignment with scientific publications. Our prior work~\cite{Schreieder2025Claim2Source} showed that formalizing claims improves retrieval performance, especially for sparse retrieval models, while modifying source documents typically introduces noise and degrades performance. Therefore, we restrict style transfer to claims and keep sources unchanged. Building on this setup, we extend the analysis to a multilingual setting and adopt LLMs (\textit{Qwen/Qwen3.5-9B} and \textit{Qwen/Qwen3.5-27B}) for zero-shot style transfer. To ensure comparability with our previous study~\cite{Schreieder2025Claim2Source}, we reuse its prompt templates. This enables a controlled comparison of model sizes while providing a more efficient and scalable setup for retrieval-oriented applications. Due to the non-deterministic behavior of LLM-based style transfer, all experiments are repeated three times and results are reported as averages across runs. We evaluate the following four claim styles:

\bslabel{Formal.} Claims are rewritten in a more formal style by removing informal elements such as hashtags and emojis while preserving the original meaning.

\bslabel{Scientific.} Claims are reformulated as scientific statements using domain-specific terminology, aiming to better match the language of scientific publications.

\bslabel{Abstract.} Claims are expanded into short abstract-like texts that summarize the underlying statement in a structured and coherent form, increasing stylistic similarity to scientific documents.

\bslabel{Question.} Claims are converted into concise scientific questions, aligning with the input format of retrieval models optimized for question-based queries.

\subsection{Similarity-based Re-Ranking}
\label{sec:methodology_similarity}

Re-ranking has proven to be an effective component in scientific claim--source retrieval, as top-performing systems in the CheckThat! 2025 shared task incorporated re-ranking approaches to improve retrieval performance \cite{Alam2026CheckThat2025}. Motivated by these findings, we analyze the impact of similarity-based re-ranking in our setting.
Similarity-based re-ranking estimates the relevance of a claim--source pair from their textual content, in contrast to the signal-based approach described in Section~\ref{sec:methodology_signal}. For all experiments, we use the top-100 sources retrieved by GritLM as candidates. The re-rankers assign new relevance scores to these candidates and generate a new ranking independently of the original retrieval scores.

\bslabel{Nemotron.} The \textit{nvidia/llama-nemotron-rerank-vl-1b-v2} model is a pointwise cross-encoder re-ranker with approximately 1.7B parameters.

\bslabel{BGE.} The \textit{BAAI/bge-reranker-v2-m3} model is a multilingual pointwise cross-encoder re-ranker with approximately 0.6B parameters from the BGE family~\cite{Chen2024M3}.

\bslabel{Jina.} The \textit{jinaai/jina-reranker-v3} model is a multilingual listwise re-ranker with approximately 0.6B parameters that jointly ranks candidate sources by considering their relative relevance within a shared context~\cite{Wang2025Jina}.

\bslabel{Qwen3.} The models \textit{Qwen/Qwen3-Reranker-0.6B} and \textit{Qwen/Qwen3-Reranker-8B} are pointwise generative re-rankers from the Qwen3 family~\cite{Zhang2025Qwen3}. Their different model sizes additionally enable comparison of scale effects in re-ranking.

\subsection{Signal-based Re-Ranking}
\label{sec:methodology_signal}

\begin{figure}[t]
    \centering
    \includegraphics[width=\linewidth]{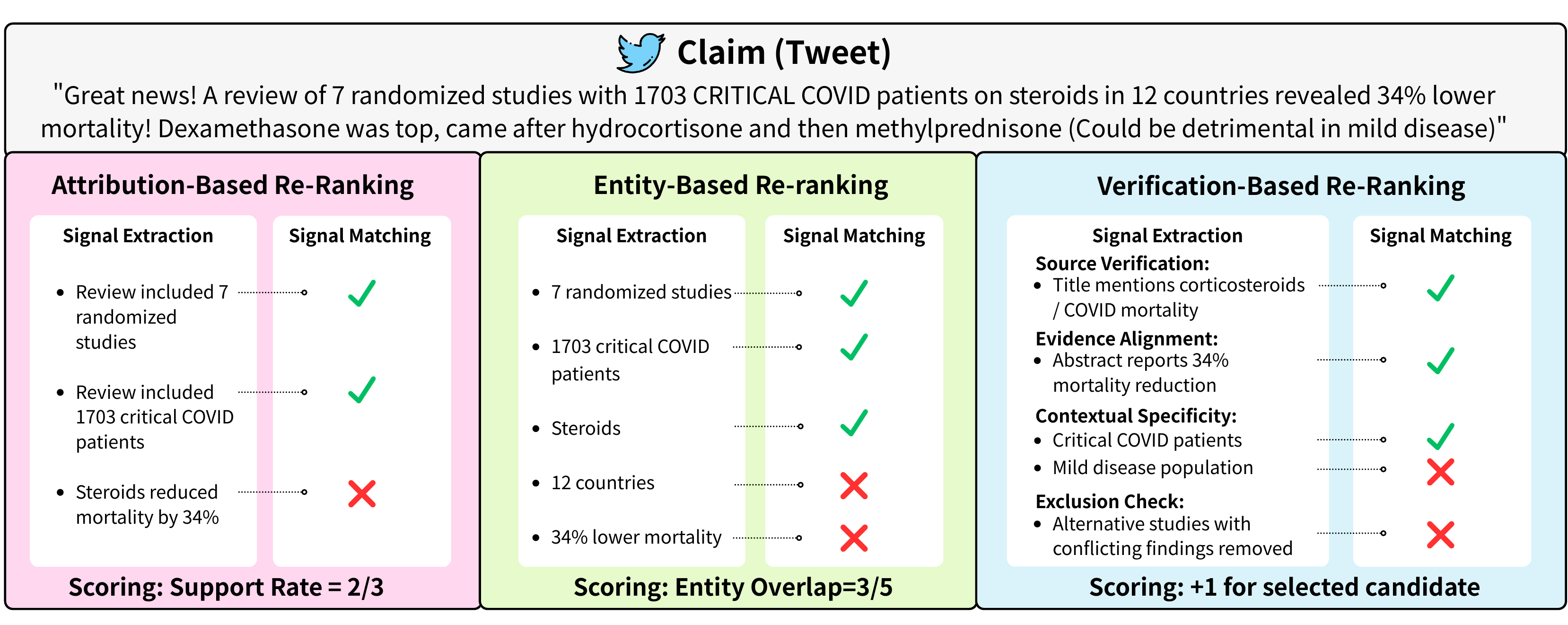}
    \caption{Overview of the signal-based re-ranking process. Signals extracted from a claim are used to re-rank top-k sources, where sources are scored according to signal matching.}
    \label{fig:signal_based_re_ranking}
\end{figure}

As illustrated in Figure~\ref{fig:signal_based_re_ranking}, initial retrieval identifies a broad set of candidate sources, while similarity-based re-ranking refines them based on semantic relevance. However, semantic similarity alone may not fully capture claim--source alignment. We therefore investigate signal-based re-ranking, which incorporates structured signals to improve retrieval performance. Due to the computational cost of LLM-based reasoning, signal-based re-ranking is applied only to the top-10 candidates from the best-performing retrieval and similarity-based re-ranking setup. We evaluate the proposed re-ranking approaches using three reasoning LLMs. All experiments are repeated three times and averaged across runs.

\bslabel{Qwen3.5.} \textit{Qwen/Qwen3.5-27B} is a 27B-parameter reasoning model from the Qwen family with strong instruction-following and multilingual capabilities.

\bslabel{Gemma-4.} \textit{google/gemma-4-31B-it} is a 31B-parameter instruction-tuned model from the Gemma family with strong reasoning and multilingual performance.

\bslabel{Kimi-2.6.} \textit{moonshotai/Kimi-K2.6} is a large-scale mixture-of-experts model with approximately 1T total parameters and 32B activated parameters during inference, enabling comparison against frontier-scale reasoning models.

We investigate the following three novel signal-based re-rankers.

\bslabel{Attribution-based Re-Ranking.} Attribution is the process of linking content to supporting evidence and represents a key component of evidence-based text generation with LLMs~\cite{Schreieder2026Survey}. While attribution approaches have been extensively studied in this context, their use as a re-ranking signal for scientific claim--source retrieval remains largely unexplored. We therefore draw inspiration from the atomic fact decomposition introduced in FactScore~\cite{min-etal-2023-factscore}. Using a zero-shot prompt, claims are decomposed into atomic facts while preserving their content and contextual information. Each atomic fact is then evaluated independently against source documents through natural language inference with the respective reasoning LLM~\cite{Schreieder2026Survey}. The resulting fact-level predictions are aggregated into a support rate representing the proportion of supported facts, which is used as an attribution signal for re-ranking candidate sources.

\bslabel{Entity-based Re-Ranking.} Salient entities often provide retrieval signals that are not fully captured by semantic similarity~\cite{Shavarani2025Entity}. In scientific claim--source retrieval, social media claims frequently contain implicit references through entities such as authors, institutions, datasets, or domain-specific terminology. We therefore analyze whether explicit entity overlap between claims and source documents provides a useful signal for claim--source alignment. Using a structured zero-shot prompt, entities are extracted and compared across claim--source pairs. The prompt considers biomedical, chemical, numerical, methodological, institutional, geographic, and temporal entities, while abbreviations and common synonyms are normalized. The re-ranking score is computed as the proportion of claim entities that also occur in the source representation.

\bslabel{Verification-based Re-Ranking.} This approach addresses cases where semantic similarity or entity overlap alone is insufficient to determine whether a source actually underlies a claim. Source documents may share similar terminology while differing in the specific evidence, population, location, or intervention described. We therefore investigate whether verification reasoning provides an additional signal for claim--source alignment. Using a structured zero-shot prompt, the model selects the single most relevant source candidate for a given claim by considering source verification, evidence alignment, contextual specificity, and conflicting findings. The selected source is assigned the highest relevance score and used for re-ranking candidate sources.

%% file: texts/5_Evaluation.tex
\section{Evaluation}
\label{sec:evaluation}

We follow a sequential evaluation strategy in which each stage builds upon the strongest results of the previous one. We first identify the most effective strategy for handling multilingual claims and subsequently evaluate the impact of leveraging publication metadata. The resulting setup forms the basis for our comparative study of style transfer and re-ranking approaches. Building on this setup, we compare four claim style transfer approaches and four similarity-based re-rankers across all retrieval models. Finally, we assess the three signal-based re-ranking approaches using the best combination of retrieval model and similarity-based re-ranker. Following the CheckThat! 2026 shared task setup, all experiments are evaluated using \textit{Mean Reciprocal Rank at }5 (MRR@5).

\subsection{Analysis of Strategies for Handling Multilingual Claims}
\label{sec:evaluation_methodology_multilinguality}

Table~\ref{tab:results_multiliguality} shows the retrieval performance of different strategies for handling multilingual claims. Across all retrieval models, translating claims into English consistently achieves the highest performance. Compared to original claims, the largest improvement is observed for E5, where the average MRR@5 increases from 0.348 to 0.515 (+0.167), closely followed by BM25 with an improvement of +0.165. GTR and GritLM also show notable gains of +0.125 and +0.041, respectively. These improvements indicate that reducing the language mismatch between non-English claims and the English sources benefits both sparse and dense retrieval models. Bilingual claims outperform the original multilingual setting but consistently remain below the translated claims. These findings indicate that combining original and translated representations may create competing signals that weaken retrieval performance compared to translated claims. Based on these findings, translated claims are selected for all subsequent experiments.

\begin{table}[t!]
\centering
\caption{Retrieval performance (MRR@5) across different strategies for handling multilingual claims, including original, translated, and bilingual claim representations.}
\label{tab:results_multiliguality}
\setlength{\tabcolsep}{2pt}

\begin{tabular}{lcccc@{\hspace{6pt}}cccc@{\hspace{6pt}}cccc}
\toprule
\textbf{Models}
& \multicolumn{4}{c}{\textbf{Original Claims}} 
& \multicolumn{4}{c}{\textbf{Translated Claims}} 
& \multicolumn{4}{c}{\textbf{Bilingual Claims}} \\
\cmidrule(lr){2-5} \cmidrule(lr){6-9} \cmidrule(lr){10-13}
& EN & DE & FR & AVG 
& EN & DE & FR & AVG 
& EN & DE & FR & AVG \\
\midrule
BM25   & 0.420 & 0.078 & 0.069 & 0.189 & 0.420 & 0.252 & 0.391 & \underline{0.354} & 0.420 & 0.162 & 0.167 & 0.249 \\
GTR    & 0.476 & 0.212 & 0.300 & 0.329 & 0.476 & 0.381 & 0.505 & \underline{0.454} & 0.476 & 0.366 & 0.481 & 0.441 \\
E5     & 0.520 & 0.187 & 0.336 & 0.348 & 0.520 & 0.468 & 0.556 & \underline{0.515} & 0.520 & 0.396 & 0.502 & 0.473 \\
GritLM & 0.648 & 0.494 & 0.624 & 0.589 & 0.648 & 0.573 & 0.668 & \underline{0.630} & 0.648 & 0.527 & 0.653 & 0.609 \\
\bottomrule
\end{tabular}
\end{table}

\subsection{Analysis of Strategies for Leveraging Publication Metadata}
\label{sec:evaluation_methodology_metadata}

Table~\ref{tab:results_metadata} summarizes the impact of leveraging publication metadata on retrieval performance. Across all retrieval models, augmenting the source representation with publication metadata consistently improves retrieval performance, although the magnitude of the gains varies between models. The largest improvement is observed for BM25, where the average MRR@5 increases from 0.354 to 0.382 (+0.028). Smaller gains are observed for GTR (+0.003), E5 (+0.004), and GritLM (+0.003). The larger gains observed for BM25 indicate that sparse retrieval benefits more from additional metadata fields. In contrast, dense retrieval models already capture semantic relationships effectively from titles and abstracts, resulting in more limited gains from leveraging metadata. Based on these findings, publication metadata is included in all subsequent experiments.

\begin{table*}[t]
\centering
\caption{Retrieval performance (MRR@5) on translated claims across different strategies for leveraging publication metadata.}
\label{tab:results_metadata}
\setlength{\tabcolsep}{3pt}

\begin{tabular}{lcccc@{\hspace{6pt}}cccc}
\toprule
\textbf{Model}
& \multicolumn{4}{c}{\textbf{No Metadata}}
& \multicolumn{4}{c}{\textbf{Metadata}} \\
\cmidrule(lr){2-5} \cmidrule(lr){6-9}
& EN & DE & FR & AVG
& EN & DE & FR & AVG \\
\midrule
BM25   & 0.420 & 0.252 & 0.391 & 0.354 & 0.432 & 0.284 & 0.431 & \underline{0.382} \\
GTR    & 0.476 & 0.381 & 0.505 & 0.454 & 0.482 & 0.381 & 0.509 & \underline{0.457} \\
E5     & 0.520 & 0.468 & 0.556 & 0.515 & 0.533 & 0.455 & 0.569 & \underline{0.519} \\
GritLM & 0.648 & 0.573 & 0.668 & 0.630 & 0.660 & 0.567 & 0.672 & \underline{0.633} \\
\bottomrule
\end{tabular}
\end{table*}

\subsection{Impact of Style Transfer}
\label{sec:evaluation_methodology_style_transfer}

Table~\ref{tab:results_style_transfer} summarizes the impact of different claim style transfer approaches on retrieval performance. The results show that the effectiveness of style transfer strongly depends on the retrieval model. For BM25, the highest performance is achieved with formal claims, improving MRR@5 from 0.382 to 0.401 (+0.019). This observation is consistent with prior findings~\cite{Schreieder2025Claim2Source}, where formalizing claims also yielded the strongest improvements for sparse retrieval. In contrast, GritLM achieves its highest performance without applying style transfer, reaching an MRR@5 of 0.633. This result suggests that stronger retrieval models may already be robust to stylistic variations in claims. For dense retrieval models, a different pattern emerges. Both GTR and E5 achieve their highest performance with question-style claims, improving from 0.457 to 0.466 (+0.009) and from 0.519 to 0.545 (+0.026), respectively. Unlike prior findings~\cite{Schreieder2025Claim2Source}, where formal and scientific claim styles showed the strongest improvements, question-based formulations consistently achieve the highest performance in the current multilingual setting. One explanation is that question formulations align more closely with retrieval objectives by expressing information needs more explicitly. Across most style transfer settings, Qwen3.5-27B slightly outperforms Qwen3.5-9B, indicating that larger LLMs produce modest improvements in claim transformations. Results were stable across runs, with standard deviations $\leq 0.004$ MRR@5. Overall, style transfer improves retrieval performance for most models, although the optimal claim style depends on the underlying retrieval architecture.

\begin{table}[h]
\centering
\caption{Retrieval performance (MRR@5) on translated claims with publication metadata for style transfer using Qwen3.5-9B and Qwen3.5-27B. Original (Orig.) denotes claims without style transfer, while Formal (Form.), Scientific (Sci.), Abstract (Abs.), and Question (Quest.) denote the evaluated claim styles. * marks a statistically significant improvement over Orig. (paired Student's $t$-test, $p < 0.05$).}
\label{tab:results_style_transfer}
\setlength{\tabcolsep}{4pt}

\begin{tabular}{lccccccccc}
\toprule
& & \multicolumn{4}{c}{\textbf{Qwen3.5-9B}} & \multicolumn{4}{c}{\textbf{Qwen3.5-27B}} \\
\cmidrule(lr){3-6} \cmidrule(lr){7-10}
\textbf{Model} & \textbf{Orig.} & \textbf{Form.} & \textbf{Sci.} & \textbf{Abs.} & \textbf{Quest.}
& \textbf{Form.} & \textbf{Sci.} & \textbf{Abs.} & \textbf{Quest.} \\
\midrule
BM25   & 0.382 & \underline{0.397}\sig & 0.394\sig & 0.298 & 0.383 
                & \underline{0.401}\sig & 0.397\sig & 0.324 & 0.385 \\

GTR    & 0.457 & 0.452 & 0.445 & 0.411 & \underline{0.458}
                & 0.451 & 0.449 & 0.420 & \underline{0.466}\sig \\

E5     & 0.519 & 0.524\sig & 0.530\sig & 0.471 & \underline{0.544}\sig
                & 0.528\sig & 0.535\sig & 0.486 & \underline{0.545}\sig \\

GritLM & \underline{0.633} & 0.629 & 0.602 & 0.567 & 0.601
                & 0.630 & 0.615 & 0.576 & 0.608 \\
\bottomrule
\end{tabular}
\end{table}

\subsection{Impact of Similarity-based Re-Ranking}
\label{sec:evaluation_methodology_similarity}

Similarity-based re-ranking improves retrieval performance for most models over the initial GritLM retrieval stage (see Table~\ref{tab:results_similarity_reranking}). The largest improvement is achieved by Qwen3-8B, increasing average MRR@5 from 0.633 to 0.733 (+0.100), with consistent gains across all languages. Nemotron and Jina also improve performance, reaching average MRR@5 scores of 0.686 (+0.053) and 0.659 (+0.026), respectively. In contrast, Qwen3-0.6B slightly underperforms the retrieval baseline (-0.008), while BGE substantially reduces performance (-0.115). The large performance difference between Qwen3-0.6B and Qwen3-8B further highlights that model capacity plays an important role in similarity-based re-ranking.

\begin{table}[t]
\centering
\caption{Retrieval performance (MRR@5) on translated claims with publication metadata using similarity-based re-ranking on top of GritLM retrieval. Baseline denotes GritLM without re-ranking, while Nemotron, BGE, Jina, Qwen3-0.6B, and Qwen3-8B denote the evaluated similarity-based re-rankers.}
\label{tab:results_similarity_reranking}
\setlength{\tabcolsep}{4pt}

\begin{tabular}{lcccc}
\toprule
\textbf{Re-Ranker} & \textbf{EN} & \textbf{DE} & \textbf{FR} & \textbf{AVG} \\
\midrule
GritLM          & 0.660 & 0.567 & 0.673 & 0.633 \\
\midrule
Nemotron        & 0.714 & 0.606 & 0.740 & 0.686 \\
BGE             & 0.542 & 0.450 & 0.563 & 0.518 \\
Jina            & 0.688 & 0.583 & 0.707 & 0.659 \\
Qwen3-0.6B      & 0.653 & 0.546 & 0.676 & 0.625 \\
Qwen3-8B        & \underline{0.745} & \underline{0.675} & \underline{0.778} & \underline{0.733} \\
\bottomrule
\end{tabular}
\end{table}

\subsection{Impact of Signal-based Re-Ranking}
\label{sec:evaluation_methodology_signal}

Signal-based re-ranking leads to different retrieval behavior depending on the signal type (see Table~\ref{tab:reranker_results}). Attribution-based re-ranking performs comparably to the GritLM+Qwen3-8B baseline, with only minor performance differences across models and languages. Kimi-2.6 achieves the highest score for attribution-based re-ranking with an average MRR@5 of 0.735 (+0.002). In contrast, entity-based re-ranking consistently reduces performance across all models, decreasing average MRR@5 by 0.020 to 0.041 relative to the baseline. Verification-based re-ranking yields the strongest improvements, with Kimi-2.6 achieving the best performance and increasing average MRR@5 from 0.733 to 0.758 (+0.025). Gemma-4 also improves performance through verification-based re-ranking, reaching an average MRR@5 of 0.745 (+0.012), while Qwen3.5 matches the baseline. Results were stable across runs, with standard deviations $\leq 0.004$.

\begin{table}[t]
\centering
\caption{Retrieval performance (MRR@5) of signal-based re-ranking on top of GritLM retrieval with Qwen3-8B similarity-based re-ranking. Baseline denotes the selected retrieval and similarity-based re-ranking setup. * marks a statistically significant improvement in AVG over the baseline (paired Student's $t$-test, $p < 0.05$).}
\label{tab:reranker_results}
\setlength{\tabcolsep}{2pt}

\begin{tabular}{lcccc@{\hspace{5pt}}cccc@{\hspace{5pt}}cccc}
\toprule
\textbf{Model}
& \multicolumn{4}{c}{\textbf{Attribution}}
& \multicolumn{4}{c}{\textbf{Entity}}
& \multicolumn{4}{c}{\textbf{Verification}} \\
\cmidrule(lr){2-5}
\cmidrule(lr){6-9}
\cmidrule(lr){10-13}

& EN & DE & FR & AVG
& EN & DE & FR & AVG
& EN & DE & FR & AVG \\
\midrule

Baseline
& 0.745 & 0.675 & 0.778 & 0.733
& 0.745 & 0.675 & 0.778 & \underline{0.733}
& 0.745 & 0.675 & 0.778 & 0.733 \\
\midrule

Qwen3.5
& 0.744 & 0.671 & 0.782 & 0.732
& 0.700 & 0.629 & 0.747 & 0.692
& 0.749 & 0.662 & 0.789 & 0.733 \\

Gemma-4
& 0.743 & 0.671 & 0.779 & 0.731
& 0.699 & 0.640 & 0.738 & 0.692
& 0.763 & 0.673 & 0.798 & 0.745\sig \\

Kimi-2.6
& 0.748 & 0.678 & 0.779 &  \underline{0.735}\sig
& 0.726 & 0.654 & 0.759 & 0.713
& 0.782 & 0.692 & 0.801 & \underline{0.758}\sig \\
\bottomrule
\end{tabular}
\end{table}

%% file: texts/6_Discussion.tex
\section{Discussion}
\label{sec:discussion}

Our findings reveal several insights into scientific claim--source retrieval.

\bslabel{Reducing Language Mismatch Improves Retrieval Performance.} Translating claims into English consistently improves retrieval performance across all tested models, with the largest gain observed for E5 (+0.167 MRR@5). While a bilingual claim representation also improves performance compared to the original multilingual setting, translated claims consistently achieve the strongest results. These findings highlight that language differences between social media claims and source publications remain a major challenge for scientific claim--source retrieval, emphasizing the importance of reducing such mismatch.

\bslabel{Structured Source Representations Improve Retrieval Performance.} Representing sources through structured publication information provides useful retrieval signals, particularly for BM25, which shows an improvement of +0.028 MRR@5. The stronger gains for sparse retrieval suggest that structured metadata can provide useful lexical cues beyond titles and abstracts. This underscores the importance of explicitly incorporating such indirect signals into source representations for scientific claim--source retrieval.

\bslabel{Style Transfer Compensates for Retrieval Limitations.} The effectiveness of style transfer appears to depend on the retrieval objective underlying the model architecture. BM25 benefits most from formal claim styles (+0.019 MRR@5), as sparse retrieval relies on exact term matching between claims and source documents. In contrast, question-style claims yield the strongest improvements for GTR (+0.009) and E5 (+0.026), suggesting that explicitly reformulating claims as information needs better aligns with the retrieval objectives of these models. Interestingly, GritLM achieves its highest performance without style transfer, while all transformed claim styles reduce retrieval effectiveness. Overall, these findings highlight that style transfer primarily compensates for retrieval limitations rather than providing universal improvements.

\bslabel{Structured Re-Ranking Signals Differ in Effectiveness.} Structured re-ranking signals differ substantially in effectiveness. Similarity-based re-ranking provides the largest gains (+0.100 MRR@5 for Qwen3-8B), while verification-based re-ranking yields further improvements (+0.025 for Kimi-2.6). In contrast, attribution-based re-ranking yields only marginal performance gains, whereas entity-based re-ranking consistently reduces retrieval performance. These findings show that not all structured signals are equally useful for scientific claim--source retrieval. In particular, verification signals can refine candidate rankings beyond semantic similarity alone. However, such improvements come at substantially higher computational cost, as LLM-based verification requires expensive reasoning even when applied only to a small set of top-ranked candidates.

%% file: texts/7_Conclusion.tex
\section{Conclusion}
\label{sec:conclusion}

In this paper, we presented a comparative study of scientific claim--source retrieval on the CheckThat! 2026 benchmark. Our results show that claim translation improves retrieval performance, while publication metadata provides additional retrieval signals beyond titles and abstracts. Furthermore, the effectiveness of style transfer depends strongly on the retrieval model and its underlying objective. Among the evaluated re-ranking approaches, similarity-based re-ranking yields the largest performance gains, while verification-based re-ranking further improves source selection by explicitly verifying claim--source alignment.

%% file: literature.bib
@inproceedings{Schreieder2026Survey,
    title = "Attribution, Citation, and Quotation: A Survey of Evidence-based Text Generation with Large Language Models",
    author = {Schreieder, Tobias  and
      Schopf, Tim  and
      F{\"a}rber, Michael},
    editor = "Liakata, Maria  and
      Moreira, Viviane P.  and
      Zhang, Jiajun  and
      Jurgens, David",
    booktitle = "Proceedings of the 64th Annual Meeting of the {A}ssociation for {C}omputational {L}inguistics (Volume 1: Long Papers)",
    month = jul,
    year = "2026",
    address = "San Diego, California, United States",
    publisher = "Association for Computational Linguistics",
    doi = "10.18653/v1/2026.acl-long.1430",
    pages = "30956--31000"
}

@InProceedings{Schreieder2025Claim2Source,
  author    = {Schreieder, Tobias and F\"{a}rber, Michael},
  title     = {Claim2Source at CheckThat! 2025: Zero-Shot Style Transfer for Scientific Claim-Source Retrieval},
  booktitle = {Working Notes of CLEF 2025 - Conference and Labs of the Evaluation Forum},
  series    = {CLEF~2025},
  address   = {Madrid, Spain},
  year      = {2025},
  url       = {https://ceur-ws.org/Vol-4038/paper_94.pdf}
}

@inproceedings{Besrour2025SQuAI,
author = {Besrour, Ines and He, Jingbo and Schreieder, Tobias and F\"{a}rber, Michael},
title = {SQuAI: Scientific Question-Answering with Multi-Agent Retrieval-Augmented Generation},
year = {2025},
isbn = {9798400720406},
publisher = {Association for Computing Machinery},
address = {New York, NY, USA},
doi = {10.1145/3746252.3761471},
booktitle = {Proceedings of the 34th ACM International Conference on Information and Knowledge Management},
pages = {6603–6608},
numpages = {6},
location = {Seoul, Republic of Korea},
series = {CIKM '25}
}

@InProceedings{Sager2025DeepRetrieval,
  author    = {Sager, Pascal J. and Kamaraj, Ashwini and Grewe, Benjamin F. and Stadelmann, Thilo},
  title     = {Deep Retrieval at CheckThat! 2025: Identifying Scientific Papers from Implicit Social Media Mentions via Hybrid Retrieval and Re-Ranking},
  booktitle = {Working Notes of CLEF 2025 - Conference and Labs of the Evaluation Forum},
  series    = {CLEF~2025},
  address   = {Madrid, Spain},
  year      = {2025},
  url       = {https://ceur-ws.org/Vol-4038/paper_89.pdf}
}

@InProceedings{Schofield025DSGT,
  author    = {Schofield, Jeanette and Tian, Shuyu and Truong, Hoang Thanh Thanh and Heil, Maximilian},
  title     = {DS@GT at CheckThat! 2025: Exploring Retrieval and Reranking Pipelines for Scientific Claim Source Retrieval on Social Media Discourse},
  booktitle = {Working Notes of CLEF 2025 - Conference and Labs of the Evaluation Forum},
  series    = {CLEF~2025},
  address   = {Madrid, Spain},
  year      = {2025},
  url       = {https://ceur-ws.org/Vol-4038/paper_93.pdf}
}

@InProceedings{Staudinger025ATOM,
  author    = {Staudinger, Moritz and El-Ebshihy, Alaa and Kusa, Wojciech and Piroi, Florina and Hanbury, Allan},
  title     = {ATOM at CheckThat! 2025: Retrieve the Implicit - Scientific Evidence Retrieval},
  booktitle = {Working Notes of CLEF 2025 - Conference and Labs of the Evaluation Forum},
  series    = {CLEF~2025},
  address   = {Madrid, Spain},
  year      = {2025},
  url       = {https://ceur-ws.org/Vol-4038/paper_98.pdf}
}

@InProceedings{Ma2018TwitterRetrieval,
author="Ma, Wenjia
and Chao, Wenhan
and Luo, Zhunchen
and Jiang, Xin",
editor="Hacid, Hakim
and Cellary, Wojciech
and Wang, Hua
and Paik, Hye-Young
and Zhou, Rui",
title="Claim Retrieval in Twitter",
booktitle="Web Information Systems Engineering -- WISE 2018",
year="2018",
publisher="Springer International Publishing",
address="Cham",
pages="297--307",
doi="10.1007/978-3-030-02922-7_20"
}

@InProceedings{Soleimani2020BERTRetrieval,
author="Soleimani, Amir
and Monz, Christof
and Worring, Marcel",
editor="Jose, Joemon M.
and Yilmaz, Emine
and Magalh{\~a}es, Jo{\~a}o
and Castells, Pablo
and Ferro, Nicola
and Silva, M{\'a}rio J.
and Martins, Fl{\'a}vio",
title="BERT for Evidence Retrieval and Claim Verification",
booktitle="Advances in Information Retrieval",
year="2020",
publisher="Springer International Publishing",
address="Cham",
pages="359--366",
doi = {10.1007/978-3-030-45442-5_45 }
}

@inproceedings{Samarinas2021AutomatedFactChecking,
    title = "Improving Evidence Retrieval for Automated Explainable Fact-Checking",
    author = "Samarinas, Chris  and
      Hsu, Wynne  and
      Lee, Mong Li",
    editor = "Sil, Avi  and
      Lin, Xi Victoria",
    booktitle = "Proceedings of the 2021 Conference of the North American Chapter of the Association for Computational Linguistics: Human Language Technologies: Demonstrations",
    month = jun,
    year = "2021",
    address = "Online",
    publisher = "Association for Computational Linguistics",
    doi = "10.18653/v1/2021.naacl-demos.10",
    pages = "84--91"
}

@inproceedings{Sundriyal2022Covid19Retrieval,
    title = "Document Retrieval and Claim Verification to Mitigate {COVID}-19 Misinformation",
    author = "Sundriyal, Megha  and
      Malhotra, Ganeshan  and
      Akhtar, Md Shad  and
      Sengupta, Shubhashis  and
      Fano, Andrew  and
      Chakraborty, Tanmoy",
    booktitle = "Proceedings of the Workshop on Combating Online Hostile Posts in Regional Languages during Emergency Situations",
    year = "2022",
    address = "Dublin, Ireland",
    publisher = "Association for Computational Linguistics",
    doi = "10.18653/v1/2022.constraint-1.8",
    pages = "66--74"
}

@InProceedings{Zuo2023CrossGenreRetrieval,
author="Zuo, Chaoyuan
and Wang, Chenlu
and Banerjee, Ritwik",
editor="Yang, Xiaochun
and Suhartanto, Heru
and Wang, Guoren
and Wang, Bin
and Jiang, Jing
and Li, Bing
and Zhu, Huaijie
and Cui, Ningning",
title="Cross-Genre Retrieval for Information Integrity: A COVID-19 Case Study",
booktitle="Advanced Data Mining and Applications",
year="2023",
publisher="Springer Nature Switzerland",
address="Cham",
pages="495--509",
doi="10.1007/978-3-031-46677-9_34"
}

@inproceedings{Zhang2023FeedbackFactVerification,
    title = "From Relevance to Utility: Evidence Retrieval with Feedback for Fact Verification",
    author = "Zhang, Hengran  and
      Zhang, Ruqing  and
      Guo, Jiafeng  and
      de Rijke, Maarten  and
      Fan, Yixing  and
      Cheng, Xueqi",
    editor = "Bouamor, Houda  and
      Pino, Juan  and
      Bali, Kalika",
    booktitle = "Findings of the Association for Computational Linguistics: EMNLP 2023",
    month = dec,
    year = "2023",
    address = "Singapore",
    publisher = "Association for Computational Linguistics",
    doi = "10.18653/v1/2023.findings-emnlp.422",
    pages = "6373--6384",
}

@inproceedings{Liao2023MUSER,
author = {Liao, Hao and Peng, Jiahao and Huang, Zhanyi and Zhang, Wei and Li, Guanghua and Shu, Kai and Xie, Xing},
title = {MUSER: A MUlti-Step Evidence Retrieval Enhancement Framework for Fake News Detection},
year = {2023},
isbn = {9798400701030},
publisher = {Association for Computing Machinery},
address = {New York, NY, USA},
doi = {10.1145/3580305.3599873},
booktitle = {Proceedings of the 29th ACM SIGKDD Conference on Knowledge Discovery and Data Mining},
pages = {4461–4472},
numpages = {12},
location = {Long Beach, CA, USA},
series = {KDD '23}
}

@inproceedings{Chen2024ComplexClaimVerification,
    title = "Complex Claim Verification with Evidence Retrieved in the Wild",
    author = "Chen, Jifan  and
      Kim, Grace  and
      Sriram, Aniruddh  and
      Durrett, Greg  and
      Choi, Eunsol",
    editor = "Duh, Kevin  and
      Gomez, Helena  and
      Bethard, Steven",
    booktitle = "Proceedings of the 2024 Conference of the North American Chapter of the Association for Computational Linguistics: Human Language Technologies (Volume 1: Long Papers)",
    month = jun,
    year = "2024",
    address = "Mexico City, Mexico",
    publisher = "Association for Computational Linguistics",
    doi = "10.18653/v1/2024.naacl-long.196",
    pages = "3569--3587",
}

@inproceedings{Sriram2024ContrastiveLearning,
    title = "Contrastive Learning to Improve Retrieval for Real-World Fact Checking",
    author = "Sriram, Aniruddh  and
      Xu, Fangyuan  and
      Choi, Eunsol  and
      Durrett, Greg",
    editor = "Schlichtkrull, Michael  and
      Chen, Yulong  and
      Whitehouse, Chenxi  and
      Deng, Zhenyun  and
      Akhtar, Mubashara  and
      Aly, Rami  and
      Guo, Zhijiang  and
      Christodoulopoulos, Christos  and
      Cocarascu, Oana  and
      Mittal, Arpit  and
      Thorne, James  and
      Vlachos, Andreas",
    booktitle = "Proceedings of the Seventh Fact Extraction and VERification Workshop (FEVER)",
    month = nov,
    year = "2024",
    address = "Miami, Florida, USA",
    publisher = "Association for Computational Linguistics",
    doi = "10.18653/v1/2024.fever-1.28",
    pages = "264--279",
}

@inproceedings{Churina2024QuestionEnrichment,
    title = "Improving Evidence Retrieval on Claim Verification Pipeline through Question Enrichment",
    author = "Churina, Svetlana  and
      Barik, Anab Maulana  and
      Phaye, Saisamarth Rajesh",
    editor = "Schlichtkrull, Michael  and
      Chen, Yulong  and
      Whitehouse, Chenxi  and
      Deng, Zhenyun  and
      Akhtar, Mubashara  and
      Aly, Rami  and
      Guo, Zhijiang  and
      Christodoulopoulos, Christos  and
      Cocarascu, Oana  and
      Mittal, Arpit  and
      Thorne, James  and
      Vlachos, Andreas",
    booktitle = "Proceedings of the Seventh Fact Extraction and VERification Workshop (FEVER)",
    month = nov,
    year = "2024",
    address = "Miami, Florida, USA",
    publisher = "Association for Computational Linguistics",
    doi = "10.18653/v1/2024.fever-1.6",
    pages = "64--70",
}

@ARTICLE{Upadhyay2025RAG,
  title    = "Enhancing Health Information Retrieval with {RAG} by prioritizing
              topical relevance and factual accuracy",
  author   = "Upadhyay, Rishabh and Viviani, Marco",
  journal  = "Discover Computing",
  volume   =  28,
  number   =  1,
  pages    = "27",
  year     =  2025,
  doi      =  "10.1007/s10791-025-09505-5"
}

@inproceedings{Shavarani2025Entity,
    title = "Entity Retrieval for Answering Entity-Centric Questions",
    author = "Shavarani, Hassan  and
      Sarkar, Anoop",
    editor = "Shi, Weijia  and
      Yu, Wenhao  and
      Asai, Akari  and
      Jiang, Meng  and
      Durrett, Greg  and
      Hajishirzi, Hannaneh  and
      Zettlemoyer, Luke",
    booktitle = "Proceedings of the 4th International Workshop on Knowledge-Augmented Methods for Natural Language Processing",
    month = may,
    year = "2025",
    address = "Albuquerque, New Mexico, USA",
    publisher = "Association for Computational Linguistics",
    doi = "10.18653/v1/2025.knowledgenlp-1.1",
    pages = "1--17"
}

@article{Haunschild2021TwitterOpioid,
    author = {Haunschild, Robin and Bornmann, Lutz and Potnis, Devendra and Tahamtan, Iman},
    title = {Investigating dissemination of scientific information on Twitter: A study of topic networks in opioid publications},
    journal = {Quantitative Science Studies},
    volume = {2},
    number = {4},
    pages = {1486-1510},
    year = {2021},
    doi = {10.1162/qss_a_00168},
}

@article{Guenther2023TwitterScienceCommunication, 
author = {Lars Guenther and Claudia Wilhelm and Corinna Oschatz and Janise Brück},
title ={Science communication on Twitter: Measuring indicators of engagement and their links to user interaction in communication scholars’ Tweet content},
journal = {Public Understanding of Science},
volume = {32},
number = {7},
pages = {860-869},
year = {2023},
doi = {10.1177/09636625231166552},
}

@article{Suarez-Lledo2021HealthMisinformation,
  title     = "Prevalence of health misinformation on social media: Systematic
               review",
  author    = "Suarez-Lledo, Victor and Alvarez-Galvez, Javier",
  journal   = "J. Med. Internet Res.",
  publisher = "JMIR Publications Inc.",
  volume    =  23,
  number    =  1,
  pages     = "e17187",
  month     =  jan,
  year      =  2021,
  language  = "en",
  doi       = "10.2196/17187"
}

@article{SwireThompson2020Misinformation,
   author = "Swire-Thompson, Briony and Lazer, David",
   title = "Public Health and Online Misinformation: Challenges and Recommendations", 
   journal= "Annual Review of Public Health",
   year = "2020",
   volume = "41",
   number = "Volume 41, 2020",
   pages = "433-451",
   doi = "10.1146/annurev-publhealth-040119-094127",
   publisher = "Annual Reviews",
   issn = "1545-2093",
   type = "Journal Article",
  }

@InProceedings{Alam2026CheckThat2025,
author="Alam, Firoj
and Stru{\ss}, Julia Maria
and Chakraborty, Tanmoy
and Dietze, Stefan
and Hafid, Salim
and Korre, Katerina
and Muti, Arianna
and Nakov, Preslav
and Ruggeri, Federico
and Schellhammer, Sebastian
and Setty, Vinay
and Sundriyal, Megha
and Todorov, Konstantin
and Venktesh, V.",
editor="Carrillo-de-Albornoz, Jorge
and Garc{\'i}a Seco de Herrera, Alba
and Gonzalo, Julio
and Plaza, Laura
and Mothe, Josiane
and Piroi, Florina
and Rosso, Paolo
and Spina, Damiano
and Faggioli, Guglielmo
and Ferro, Nicola",
title="Overview of the CLEF-2025 CheckThat! Lab: Subjectivity, Fact-Checking, Claim Normalization, and Retrieval",
booktitle="Experimental IR Meets Multilinguality, Multimodality, and Interaction",
year="2026",
publisher="Springer Nature Switzerland",
address="Cham",
pages="199--223",
doi="10.1007/978-3-032-04354-2_13"
}

@InProceedings{Struss2026CheckThat2026,
author="Stru{\ss}, Julia Maria
and Schellhammer, Sebastian
and Dietze, Stefan
and V., Venktesh
and Setty, Vinay
and Chakraborty, Tanmoy
and Nakov, Preslav
and Anand, Avishek
and Chungkham, Primakov
and Hafid, Salim
and Sahnan, Dhruv
and Todorov, Konstantin",
editor="Campos, Ricardo
and Jatowt, Adam
and Lan, Yanyan
and Aliannejadi, Mohammad
and Bauer, Christine
and MacAvaney, Sean
and Anand, Avishek
and Ren, Zhaochun
and Verberne, Suzan
and Bai, Nan
and Mansoury, Masoud",
title="The CLEF-2026 CheckThat! Lab: Advancing Multilingual Fact-Checking",
booktitle="Advances in Information Retrieval",
year="2026",
publisher="Springer Nature Switzerland",
address="Cham",
pages="325--335",
doi="10.1007/978-3-032-21321-1_43"
}

@inproceedings{Robertson1994BM25,
author = {Robertson, S. E. and Walker, S.},
title = {Some simple effective approximations to the 2-Poisson model for probabilistic weighted retrieval},
year = {1994},
doi = {10.5555/188490.188561},
publisher="Springer London",
address="London",
booktitle="SIGIR '94",
pages = {232–241},
numpages = {10},
location = {Dublin, Ireland},
}

@inproceedings{Robertson1994Okapi,
  author    = {Stephen E. Robertson and
               Steve Walker and
               Susan Jones and
               Micheline Hancock{-}Beaulieu and
               Mike Gatford},
  editor    = {Donna K. Harman},
  title     = {Okapi at {TREC-3}},
  booktitle = {Proceedings of The Third Text REtrieval Conference, {TREC} 1994, Gaithersburg,
               Maryland, USA, November 2-4, 1994},
  series    = {{NIST} Special Publication},
  volume    = {500-225},
  pages     = {109--126},
  publisher = {National Institute of Standards and Technology {(NIST)}},
  year      = {1994},
  url       = {http://trec.nist.gov/pubs/trec3/papers/city.ps.gz},
}

@misc{Wang2024E5,
      title={Text Embeddings by Weakly-Supervised Contrastive Pre-training}, 
      author={Liang Wang and Nan Yang and Xiaolong Huang and Binxing Jiao and Linjun Yang and Daxin Jiang and Rangan Majumder and Furu Wei},
      year={2024},
      eprint={2212.03533},
      archivePrefix={arXiv},
      primaryClass={cs.CL},
      url={https://arxiv.org/abs/2212.03533}, 
}

@inproceedings{Ni2022GTR,
    title = "Large Dual Encoders Are Generalizable Retrievers",
    author = "Ni, Jianmo  and
      Qu, Chen  and
      Lu, Jing  and
      Dai, Zhuyun  and
      Hernandez Abrego, Gustavo  and
      Ma, Ji  and
      Zhao, Vincent  and
      Luan, Yi  and
      Hall, Keith  and
      Chang, Ming-Wei  and
      Yang, Yinfei",
    editor = "Goldberg, Yoav  and
      Kozareva, Zornitsa  and
      Zhang, Yue",
    booktitle = "Proceedings of the 2022 Conference on Empirical Methods in Natural Language Processing",
    month = dec,
    year = "2022",
    address = "Abu Dhabi, United Arab Emirates",
    publisher = "Association for Computational Linguistics",
    doi = "10.18653/v1/2022.emnlp-main.669",
    pages = "9844--9855",
}

@inproceedings{Muennighoff2025GritLM,
title={Generative Representational Instruction Tuning},
author={Niklas Muennighoff and Hongjin SU and Liang Wang and Nan Yang and Furu Wei and Tao Yu and Amanpreet Singh and Douwe Kiela},
booktitle={The Thirteenth International Conference on Learning Representations},
year={2025},
url={https://openreview.net/forum?id=BC4lIvfSzv}
}

@misc{Wang2025Jina,
      title={jina-reranker-v3: Last but Not Late Interaction for Listwise Document Reranking}, 
      author={Feng Wang and Yuqing Li and Han Xiao},
      year={2025},
      eprint={2509.25085},
      archivePrefix={arXiv},
      primaryClass={cs.CL},
      url={https://arxiv.org/abs/2509.25085}, 
}

@misc{Zhang2025Qwen3,
      title={Qwen3 Embedding: Advancing Text Embedding and Reranking Through Foundation Models}, 
      author={Yanzhao Zhang and Mingxin Li and Dingkun Long and Xin Zhang and Huan Lin and Baosong Yang and Pengjun Xie and An Yang and Dayiheng Liu and Junyang Lin and Fei Huang and Jingren Zhou},
      year={2025},
      eprint={2506.05176},
      archivePrefix={arXiv},
      primaryClass={cs.CL},
      url={https://arxiv.org/abs/2506.05176}, 
}

@inproceedings{min-etal-2023-factscore,
    title = "{FA}ct{S}core: Fine-grained Atomic Evaluation of Factual Precision in Long Form Text Generation",
    author = "Min, Sewon  and
      Krishna, Kalpesh  and
      Lyu, Xinxi  and
      Lewis, Mike  and
      Yih, Wen-tau  and
      Koh, Pang  and
      Iyyer, Mohit  and
      Zettlemoyer, Luke  and
      Hajishirzi, Hannaneh",
    editor = "Bouamor, Houda  and
      Pino, Juan  and
      Bali, Kalika",
    booktitle = "Proceedings of the 2023 Conference on Empirical Methods in Natural Language Processing",
    month = dec,
    year = "2023",
    address = "Singapore",
    publisher = "Association for Computational Linguistics",
    doi = "10.18653/v1/2023.emnlp-main.741",
    pages = "12076--12100",
}

@inproceedings{Chen2024M3,
    title = "{M}3-Embedding: Multi-Linguality, Multi-Functionality, Multi-Granularity Text Embeddings Through Self-Knowledge Distillation",
    author = "Chen, Jianlyu  and
      Xiao, Shitao  and
      Zhang, Peitian  and
      Luo, Kun  and
      Lian, Defu  and
      Liu, Zheng",
    editor = "Ku, Lun-Wei  and
      Martins, Andre  and
      Srikumar, Vivek",
    booktitle = "Findings of the Association for Computational Linguistics: ACL 2024",
    month = aug,
    year = "2024",
    address = "Bangkok, Thailand",
    publisher = "Association for Computational Linguistics",
    doi = "10.18653/v1/2024.findings-acl.137",
    pages = "2318--2335"
}
